\documentclass{PoS}
\def\sss{\scriptscriptstyle}
\def\^#1{^{\sss #1}}
\def\_#1{_{\sss #1}}
\def\ten#1#2{^{\sss#1}_{\sss#2}}
\def\av#1{\langle #1\rangle}
\def\beq {\begin{equation}}
\def\eeqno#1{\label{#1}\end{equation}}

\def\az{a_0}
\def\baz{\bar a_0}
\def\l0{\ell_{0}}

\def\rar{\rightarrow}
\def\s{\sigma}
\def\r{\rho}
\def\l{\lambda}

\def\f{\phi}
\def\grad{\vec\nabla}
\def\div{\vec \nabla\cdot}
\def\gf{\grad\phi}

\def\vr{ \textbf{r}}

\def\beq {\begin{equation}}
\def\eeqno#1{\label{#1}\end{equation}}

\def\cmss{{\rm cm~s^{-2}}}
\def\kpc{{\rm Kpc}}

\def\az{a_{0}}

\def\l0{\ell_{0}}

\def\rar{\rightarrow}
\def\s{\sigma}
\def\l{\lambda}

\def\f{\phi}

\def\r{\rho}

\def\m{\mu}

\def\n{\nu}
\def\Up{\Upsilon}
\def\C{\Gamma}

\def\d{\delta}

\def\a{\alpha}
\def\b{\beta}
\def\c{\gamma}
\def\d{\delta}

\def\vr{{\bf r}}

\def\grad{\vec\nabla}
\def\div{\vec \nabla\cdot}
\def\gf{\grad\phi}
\def\fpg{4\pi G}

\def\gmn{g_{\m\n}}
\def\Gmn{g^{\mu \nu}}
\def\gab{g_{\alpha\beta}}

\def\hGmn{\hat \Gmn}

\def\hgmn{\hat g_{\m\n}}
\def\hgh{\hat g^{1/2}}
\def\gh{g^{1/2}}

\def\hC{\hat\C}
\def\cd#1{{}_{;#1}}
\def\mcd#1{{}_{:#1}}

\def\T{\mathcal{T}}
\def\F{\mathcal{F}}
\def\K{\mathcal{K}}
\def\L{\mathcal{L}}
\def\Tmn{\T_{\m\n}}
\def\Th{\hat{\mathcal{T}}}
\def\hTmn{\Th_{\m\n}}
\def\emn{\eta_{\m\n}}

\def\T{\mathcal{T}}
\def\Tmn{\T_{\m\n}}
\def\Th{\hat{\mathcal{T}}}
\def\hTmn{\Th_{\m\n}}

\def\M{\mathcal{M}}

\def\ten#1#2{^{#1}_{#2}}
\title{MD or DM? Modified dynamics at low accelerations vs dark matter}

\ShortTitle{New physics at low accelerations}

\author{\speaker{Mordehai Milgrom}%
         \\
        Weizmann Institute\\
        E-mail: \email{moti.milgrom@weizmann.ac.il}}

\abstract{The MOND paradigm posits a departure from standard
Newtonian dynamics, and from General Relativity, in the limit of
small accelerations. The resulting modified dynamics aim to account
for the mass discrepancies in the universe without non-baryonic dark
matter. I briefly review this paradigm with its basic tenets, and
its underlying theories--nonrelativistic and relativistic--including
a novel, bimetric MOND gravity theory. I also comment on MOND's
possible connection to, and origin in, the cosmological state of the
universe at large. Some of its main predictions, achievements, and
remaining desiderata are listed. I then succinctly pit MOND against
the competing paradigm of standard dynamics with cold, dark matter.

Some of the complaints leveled at MOND are: (i) ``MOND was {\it
designed} to fit rotation curves; so no wonder it is so successful
in predicting them''. This is both incorrect and quibbling: The
first ever MOND rotation curve analysis was undertaken more then
four years after the advent of MOND. And, even if MOND, epitomized
by a very simple formula, could have been designed to predict
hundreds of rotation curves, it would still be a great achievement.
(ii) ``MOND outperforms CDM only on small, galactic scales, where
formation physics is anyhow very messy, but falls behind in accounting for
`simpler', large-scale phenomena''. Quite contrarily, all the
salient MOND predictions on galactic scales follow as unavoidable, simple,
and immediate corollaries of the theory--{\it independent of any
messy formation scenario}--just as Kepler's laws, obeyed by all
planetary systems, follow from an underlying theory, not from
complex formation scenarios. To think, as dark-matter advocates say they do,
that the universal MOND regularities
exhibited by galaxies will one day be shown to somehow follow from
complex formation processes, is, to my mind, a delusion.
What is left for MOND to explain on
large scales is a little in comparison, and has to await a full
fledged relativistic MOND theory. (iii) ``The `bullet cluster' shows
that MOND still requires some matter that is dark''. Yes, it has
long been known that MOND does not fully remove the mass discrepancy
in the cores of galaxy clusters. Some additional still-dark matter
is needed. But this need not be THE ``dark matter''; a small amount
of the still-missing baryons, in some dark form (dead stars? cold
gas clouds?), or perhaps (sterile?) neutrinos, could fit the bill.}

\FullConference{Quarks, Strings and the Cosmos
- H\'{e}ctor Rubinstein Memorial Symposium\\
         August 9-11 ~2010\\
         AlbaNova, Stockholm, Sweden}

\begin{document}
\section{MOND introduced}
MOND, or modified dynamics, was propounded in the early 1980s
\cite{milgrom83}, as a new theory of dynamics that accounts for the
mass discrepancies in galactic systems without non-baryonic dark
matter (DM).  MOND hinges on the fact that the typical accelerations
in such systems are many orders of magnitude smaller than those
encountered in the solar system. It may be viewed, in some sense, as
a modification of standard dynamics at ``low accelerations''. Some
reviews of MOND can be found in
\cite{sm02,bek06,milgrom08,milgrom10}. A historical account of DM
and the MOND paradigms can be found in \cite{sanders10}.
\par
I start with some basic axioms for the new dynamics, such axioms as
was the constancy of the speed light in special relativity, the
equivalence of acceleration and gravity in General Relativity (GR),
or the parceling of the radiation field in quantum theory. The
reason for starting with such axioms, and not concentrating on a
specific theory right from the outset, is twofold: 1. We are not yet
in the state where we have one universally accepted theory for MOND,
even in the nonrelativistic (NR) regime. So, a set of basic
principles from which to start can serve as a useful anchor and
guide. 2. It turns out that there exists, in fact, a set of simply
formulated basic tenets, from which alone follow a large number of
major predictions, encompassing most of the basic, NR MOND
phenomenology. These predictions are thus shared by all theories
built on the axioms, which is a very useful knowledge.
\par
The basic tenets of MOND, as they apply to NR matter systems
governed by gravity, such as the solar system or galactic systems,
are: (i) Admit into physics a new constant, $\az$, with the
dimensions of acceleration. It defines the boundary between the
applicability domain of the standard dynamics (from which it
disappears) and the new dynamics (in the phenomenology of which
$\az$ appears extensively), qualitatively similar to how $c$ appears
in relativity, or $\hbar$ in quantum physics. (ii) A
``correspondence principle'': a NR MOND theory has Newtonian
dynamics as its formal limit when $\az\rar 0$. (iii) In the
opposite, deep-MOND limit, formally defined by $\az\rar \infty$,
while all masses $m\_i\rar 0$ (or $G\rar 0$), such that $m\_i\az$
(or $G\az$) remain fixed, we require the theory to become scale
invariant: dynamical evolution is invariant under
$(t,\vr)\rar\l(t,\vr)$.
\par
In more general contexts--such as in the relativistic regime, or
when non-gravitational interactions are involved--we may have to
generalize the above basic tenets. For example, we shall see that
there are relativistic MOND theories that do not tend exactly to GR
in the limit $\az\rar 0$ (while their NR limit does tend to
Newtonian gravity in the limit).
\section{\label{theories} MOND theories}
Here, I discuss briefly theories that incorporate the basic tenets
of MOND. In this connection, note the numerical coincidence between
the value of $\az\approx 10\^{-8}\cmss$, as determined from numerous
phenomena in which it appears (see below), and the values of some
cosmological acceleration parameters, as has been noted and
elaborated on many times along MOND's history (e.g. in
\cite{milgrom83,milgrom89,milgrom94,sanders98,milgrom99,blt09}). We
have
 \beq \baz\equiv2\pi\az\approx cH_0\approx (\Lambda/3)^{1/2},
 \eeqno{baza}
where $H_0$ is the Hubble constant, and $\Lambda$ the cosmological
constant (CC) as now deduced from various cosmological data. This
striking proximity, if it is more than a mere coincidence, may be an
important hint in constructing MOND theories. It would be desirable
(if not necessary) that a MOND theory should account for this
numerical proximity.
\par
Quantum theory and relativity depart from Newtonian dynamics, in the
first place, by introducing novel world pictures. At the more
``technical'' level, these novel concepts then form the basis for
specific theories described by this or that action. There are
intriguing hints that MOND also rests on some novel world picture
(e.g. \cite{milgrom99,milgrom09a}). However, so far, the only
palpable and ``practical'' progress made has been in constructing
modified actions for MOND theories, based on the standard concepts
of Newtonian dynamics and GR. These theories invariably involve some
function, put in by hand, that interpolates between the MOND and the
neoclassical (pre-MOND) regime. This, to my mind, testifies strongly
to the ``effective'' nature of these theories.
\par
The more tractable approach for building MOND theories has, so far,
proven to be modifying only the field equation for the gravitational
potential (the Poisson equation), leaving Newton's law of inertia
intact. In the relativistic context, the equivalent approach is to
adhere to metric theories, keep the standard matter action of GR,
with the standard (minimal) matter coupling to the metric, but
modify the equations that determine the metric from its matter
sources. I call such theories modified-gravity (MG) theories.
\par
Efforts in this direction have already yielded some NR theories, and
three classes of relativistic MOND theories (TeVeS, MOND adaptations
of Einstein-Aether theories, and BIMOND, to be described below).
\par
Other theories, loosely called ``modified inertia'' (MI) theories,
involve modifying Newton's second law, or generally, modifying the
matter actions (both NR and relativistic). This approach has proven
more difficult to implement, and remains relatively unexplored. It
seems to me, however, to be an attractive avenue, worth pursuing; it
is a natural framework for connecting MOND with cosmology, as
perhaps pointed to by the above coincidence. Some more detailed
results on this approach can be found in
\cite{milgrom94,milgrom99,milgrom06}.
\par
The new physics entailed by MOND shows up in full force only in
galactic systems, whose dynamics is governed by gravity. It is thus
difficult to distinguish observationally between the two classes.
Important differences in the predictions of the two classes of
theories exist even for purely gravitational systems, but they have,
so far, not been brought to bear on the question of which of the two
approaches is phenomenologically superior to the other.
\par
I now describe briefly some of the MOND theories propounded to date.
\subsection{Nonrelativistic theories}
\subsubsection{Nonlinear Poisson equation for the
gravitational potential $\f$} For a gravitating system made of
point-like, individually-test masses, whose trajectories are
$\vr\_i(t)$, we have \cite{bm84} \beq
\ddot\vr\_i=-\gf(\vr\_i),~~~~~~~~~~~~~~~
\div[\m(|\gf|/\az)\gf]=\fpg\r,\eeqno{xii} derived from a generalized
Poisson action. In the limit $\az\rar 0$: $\m(x\rar\infty)\rar 1$.
In the deep-MOND limit ($\az\rar\infty,~\az G$ constant): the first
eq.(\ref{xii}) implies that $\f$ has zero dimensions under scaling,
and thus for the second eq.(\ref{xii}) to give scale invariance we
have to have \beq \mu(x\ll 1)\propto x, \eeqno{i} and we normalize
$\az$ so equality holds. The nonlinear Poisson equation for the
gravitational potential $\f$, thus reduces to the 3-harmonic Poisson
equation $\div(|\gf|\gf)= \fpg\az\r$.
\par
The theory then becomes invariant to conformal transformations in
space, with many applications and possible matter-of-principle
ramifications \cite{milgrom97,milgrom09a}.

\subsubsection{Quasilinear formulation (QUMOND)}
The field equations for the gravitational potential (derived from an
action) are here \cite{milgrom10c} \beq \Delta\f=\div[\nu(|\gf^*|/
\az)\gf^*],~~~~ {\rm where}~~~~~~~~\Delta\f^*=\fpg\r. \eeqno{ii}
This requires  solving only a linear Poisson equation (twice), and
is rather easier to solve than the nonlinear Poisson equation. Scale
invariance in the deep-MOND limit dictates $y^{1/2}\nu(y\rar 0)\rar
const.$ (again $\az$ normalized so that the limit is 1).
\subsubsection{``Modified inertia''}
In \cite{milgrom94}, I considered (NR) action-based theories,
whereby the gravitational potential is still determined from the
Poisson equation, but the particle equation of motion is of the form
 \beq\textbf{A}[\{\vr(t)\},\az]=-\gf,
  \eeqno{huta}
(replacing $\ddot\vr=-\gf$), where $\textbf{A}$ is a functional of
the whole trajectory $\{\vr(t)\}$, with the dimensions of
acceleration. For $\az\rar 0$, $\textbf{A}\rar \ddot\vr$. Here, $\f$
has dimension $-1$ under scaling (from the Poisson equation); so, to
have scale invariance in the deep-MOND limit, we have to have
 \beq \textbf{A}[\{\vr(t)\},\az]\rar
  \az^{-1}\textbf{Q}(\{\vr(t)\}),\eeqno{v}
where $\textbf{Q}$  has dimensions of acceleration squared.
\par
I showed that if such an equation of motion is to follow from an
action principle, enjoy Galilei invariance, and have the above
Newtonian and MOND limits, it has to be time nonlocal.\footnote {An
interesting possibility, which I have explored only superficially,
is of a local theory that enjoys a symmetry other than Galilei,
which reduces to Galilei invariance for $\az \rar 0$.} Such nonlocal
theories (which also have to be nonlinear--an inherent property of
MOND theories) are not easy to construct. We do not yet have a fully
acceptable theory in this vein, even in the NR regime. Only some toy
theories have been partly explored
\cite{milgrom94,milgrom99,milgrom06}. In particular, in
\cite{milgrom99}, I discussed a heuristic idea, whereby MOND, indeed
inertia itself, can result from an effect of the vacuum, and where
the origin of $\az$ in cosmology also emerges, as per
eq.(\ref{baza}). The vacuum then serves as an absolute inertial
frame (acceleration with respect to the vacuum is detectable, e.g.,
through the Unruh effect).
\par
There is an important and robust prediction shared by all theories
in the class: For circular trajectories in an axisymmetric
potential, eq.(\ref{huta}) has to take the form
 \beq \m(V^2/R\az)V^2/R=-d\f/dR,  \eeqno{rc}
where, $V$ and $R$ are the orbital speed and radius, and $\m(x)$ is
universal for the theory, and is derived from the expression of the
action specialized to circular trajectories; $\m(x\ll 1)\approx x$,
$\m(x\gg 1)\approx 1$. Thus even without specifying the theory, this
class makes a very useful prediction for galaxy rotation curves
(RC). It is, indeed, eq.(\ref{rc})--which gives the MOND RC in
simple terms of the Newtonian curve--that has been used in all MOND
RC analyses to date. The predictions of modified-gravity theories
are less straightforward to compute, and require a computation of
the whole gravitational field for each galaxy anew.

\subsection{Relativistic theories}
\subsubsection{TeVeS}
The Tensor-Vector-Scalar (TeVeS), relativistic formulation of MOND,
has been put forth by Bekenstein \cite{bek04}, building on ideas by
Sanders \cite{sanders97}. The theory has been discussed and reviewed
extensively (e.g., recently, in \cite{skordis09,fs09}); so I
describe it very briefly.
\par
Gravity is carried by a metric $\gab$, a vector field  ${\cal
U}_\a$, and a scalar field $\f$, while matter degrees of freedom
couple minimally to the ``physical'' metric $\tilde \gab
=e^{-2\f}(\gab + {\cal U}_\a {\cal U}_\b) - e^{2\f} {\cal U}_\a
{\cal U}_\b$.
\par
TeVeS reproduces MOND phenomenology for galactic systems in the NR
limit, with a certain combination of its constants playing the role
of $\az$. In particular, when $\az\rar 0$, the NR limit goes to
Newtonian gravity. However, the relativistic theory does not go
exactly to GR in the same limit. This departure of TeVeS from GR
even for very high accelerations might leave observable effects in
the solar system (see, e.g., \cite{sagi9}).
\par
As in GR, the potential that appears in the expression for lensing
by NR masses (such as galactic systems) is the same as that which
governs the motion of massive particles.
\par
Cosmology, the CMB, and structure formation in TeVeS have been
considered in
\cite{dodelson06,skordisetal06,skordis06,skordis08,zlosnik08}. It
was shown that there are elements in TeVeS that could mimic
cosmological DM, although no fully satisfactory application of TeVeS
to cosmology has been demonstrated.
\par
Gravitational waves in TeVeS have been considered in
\cite{bek04,sagi10}.
\subsubsection{MOND versions of Einstein-Aether
theories} Zlosnik et al. \cite{zlosnik07} have proposed and
discussed MOND adaptations of Einstein-Aether (EA) theories. In EA
theories (e.g., \cite{jm01}, and references therein), gravity is
carried by a metric, $g_{\m\n}$, as well as a vector field, $A^\a$.
To the standard Einstein-Hilbert Lagrangian for the metric one adds
the terms \beq \L(A,g)=\frac{\az^2}{16\pi G}\F(\K)+
\L_L,\eeqno{lagrangeII} where
 \beq
\K=\az^{-2}\K^{\a\b}_{\c\s}A^{\c}\cd{\a}A^{\s}\cd{\b}.\eeqno{KKK}
$$\K^{\a\b}_{\c\s}=c_1g^{\a\b}g_{\c\s}+c_2\d^\a_\c\d^\b_\s
+c_3\d^\a_\s\d^\b_\c+c_4A^\a A^\b g_{\c\s},$$ and $\L_L$ is a
Lagrange multiplier term that forces the vector to be of unit
length. The asymptotic behavior of $\F$, at small and large
arguments, give the deep-MOND behavior and GR, respectively.
\subsubsection{Bimetric MOND gravity}
Bimetric MOND gravity (BIMOND)
\cite{milgrom09,milgrom10a,milgrom10b} is a class of relativistic
theories governed by the action
 \beq I=-\frac{1}{16\pi G}\int[\b\gh R
+ \a\hgh \hat R
 -2(g\hat g)^{1/4}\az\^2\M] d^4x
+I\_M(\gmn,\psi_i)+\hat I\_M(\hat g\_{\m\n},\hat\psi_i).
\eeqno{mushpa} It involves two metrics, $\gmn$ and $\hgmn$, whose
Ricci scalars are $R$ and $\hat R$ ($c=1$). $\M$ is a dimensionless,
scalar function of the two metrics and their first derivatives.
\par
The novelty in BIMOND--compared with earlier bimetric theories, much
discussed since the early 1970s--is in the choice of the interaction
term. The difference of the two Levi-Civita connections
 \beq C\ten{\a}{\b\c}=\C\ten{\a}{\b\c}-\hC\ten{\a}{\b\c},
  \eeqno{veyo}
is a tensor that acts like the relative gravitational accelerations
of the two sectors. This is particularly germane in the context of
MOND, where, with  $\az$ at our disposal, we can construct from
$\az\^{-1}C\ten{\a}{\b\c}$ dimensionless scalars to serve as
variables on which $\M$ depends. In particular, the scalars
constructed from the quadratic tensor
 \beq \Up_{\m\n}\equiv  C\ten{\c}{\m\l}C\ten{\l}{\n\c}
-C\ten{\c}{\m\n}C\ten{\l}{\l\c},  \eeqno{mamash} have particular
appeal; e.g.,
 \beq \Up=\Gmn\Up_{\m\n},~~~~~\hat\Up= \hGmn\Up_{\m\n}.
 \eeqno{papash}

\par
The terms $I\_M$ and $\hat I\_M$ are the matter actions for standard
matter and for putative twin matter (TM), whose existence is
suggested (but not required) by the double metric nature of the
theory. Matter degrees of freedom $\psi\_i$ couple only to the
standard metric $\gmn$, while TM couples only to $\hgmn$. In
matter-TM symmetric versions of the theory, assumed below for
concreteness, we have $\a=\b$, and $\gmn,~\hgmn$ appear
symmetrically in $\M$. There are good reasons to take $\b\approx 1$.
\par
Matter equations of motion in the two sectors, are the standard
ones, each with its own metric.
\par
The gravitational field equations are
 \beq  \b G\_{\m\n}+S\_{\m\n}=-8\pi G \Tmn,
~~~~~~~~~~~~ \b\hat G\_{\m\n}+\hat S\_{\m\n}=-8\pi G \hTmn,
\eeqno{nuvec} with $G\_{\m\n},~~
 \hat G\_{\m\n}$ the respective Einstein tensors, and $\Tmn$,
 $\hTmn$  the
energy-momentum tensors for the two sectors.
\par
The tensors $S\_{\m\n}$ and $\hat S\_{\m\n}$ have the schematic
forms
 \beq S_{\m\n}=(Q\{C\}\ten{\l}{\m\n})\cd{\l}
+N\{C^2\}_{\m\n}+ \az^2 P\gmn,~~~~~~~\hat S_{\m\n}=(\hat
Q\{C\}\ten{\l}{\m\n})\mcd{\l} +\hat N\{C^2\}_{\m\n}+ \az^2 \hat
P\hgmn, \eeqno{gumshata} where $\{C\}$ are tensors linear, and
$\{C^2\}$ quadratic, in $C\ten{\l}{\m\n}$.

Here are, briefly, some consequences expanded on in
 \cite{milgrom09,milgrom10a,milgrom10b}, assuming
$\b=\a=1$.

Imposing $\M'(z\rar\infty)\rar 0$, the limit of the theory for
$\az\rar 0$, is two uncoupled copies of GR with a CC, $\Lambda\sim
\az^2\M(\infty)$.

For the choice of scalar arguments as in eq.(\ref{papash}), we get
in the NR limit $\gmn\approx \emn-2\f\d_{\m\n}$, as in GR, but here
$\f$ is determined from a NR MOND theory. Thus, massive test
particles and photons ``see'' the same potential.

BIMOND, with its two metrics, naturally accommodates TM. Matter and
TM do not interact at all in the high-acceleration regime, and repel
each other in the MOND regime \cite{milgrom10a}: In the deep-MOND
regime, TM behaves as having a gravitational mass opposite in sign
to that of matter. This fits naturally with the (space) conformal
invariance of the theory in this limit.

Exactly symmetric matter-TM systems obey GR with a CC, $\Lambda\sim
\az^2\M(0)$.

Cosmology is preliminarily discussed in
\cite{cz10,milgrom09,milgrom10b}: In an interesting special case,
where the universe is matter-TM symmetric, there are no MOND
effects, apart for the appearance of the CC, $\sim \az^2$. The
``coincidence'' (\ref{baza}) thus follows naturally in BIMOND. In
this case cosmology is the standard FRW one, with a CC, retaining
the successes of early-universe cosmology, such as nucleosynthesis.
Other MOND effects, cosmological and local, only appear with
matter-TM separation (or if there is no TM). The expected CMB
fluctuations in BIMOND have not been studied yet.
\par
Structure formation: some aspects, relevant to weak-inhomogeneities,
are discussed in \cite{milgrom10b}. Even in a symmetric universe, if
the initial quantum fluctuations are not identical in the two
sectors, matter and TM segregate efficiently, since density
differences grow much faster that the sum. The inhomogeneities of
the two matter types develop, eventually, into mutually avoiding,
cosmic webs.

$S_{\m\n}$ in eqs.(\ref{nuvec})(\ref{gumshata}) can act as the
energy-momentum tensor of cosmological DM, beside its contribution
to the CC.

Note that the MOND limit occurs, in BIMOND, when the acceleration
{\it difference} in the two sectors is small compared with $\az$.
This quantity nearly equals the matter acceleration alone, in
pure-matter systems--such as presumably are the well scrutinized
systems (stars, galaxies, etc.)--but not in well mixed matter-TM
configurations.
\subsection{Theories with microscopic basis}
There are suggestions to obtain MOND phenomenology in various
microscopic-physics scenarios. For example, an omnipresent,
polarizable, dark-mater medium \cite{blt09}, an assumption of
properly tailored baryon-DM interactions \cite{bruneton09}, the
``dark fluid'' approach \cite{zhao08}, various versions of gravity
as an entropic effect \cite{lc10,ho10,kt10}, vacuum effects
\cite{milgrom99}, and Ho\v{r}ava gravity \cite{rom10}.

\section{\label{phenomenology}MOND phenomenology}
\subsection{\label{kepler}Kepler-like laws}
Clearly, each MOND theory has its own specific predictions regarding
the dynamics of gravitating systems, such as planetary systems,
galaxies, galactic systems, and the universe at large. It is
possible, however, to distill from these a number of predicted,
general laws. This is analogous to the extraction of Kepler's laws
from Newtonian dynamics. What is more, many of these corollaries
follow essentially only from the basic tenets of MOND, and are thus
valid in all MOND theories. It is also noteworthy that these MOND
laws are independent from the DM point of view, in the sense that
one can construct, for example, model populations of baryons-plus-DM
galaxies that satisfy any subset of these corollaries, but not the
others. This means that in the framework of the DM paradigm they
would each require a separate explanation. Following are a few of
these corollaries.
\begin{itemize}
\item
Asymptotic independence of orbital velocities on the size of the
orbit: Under space-time scaling, radii scale, but velocities do not,
resulting in radius-independent orbital velocities in the MOND
regime (large radii) around a point mass, $M$, (or around any mass
well within the orbital radius). This fact underlies the prediction
of asymptotically flat RCs of disc galaxies:  $V(r)\rar V_{\infty}$.
See Figs. \ref{fig2} and \ref{fig3}.
\item The mass-asymptotic-velocity relation: With only $\az$
available as a new dimensioned constant, we must have (equality,
rather than proportionality, defines the normalization of $\az$)
 \beq V_{\infty}^4=MG\az. \eeqno{mvr}
 This prediction underlies the observed baryonic Tully-Fisher relation (see Fig.
\ref{fig1}). The data is consistent with zero intrinsic scatter
about the MOND prediction \cite{mcgaugh10}.
\begin{figure}
\begin{center}
\includegraphics[width=0.5\columnwidth]{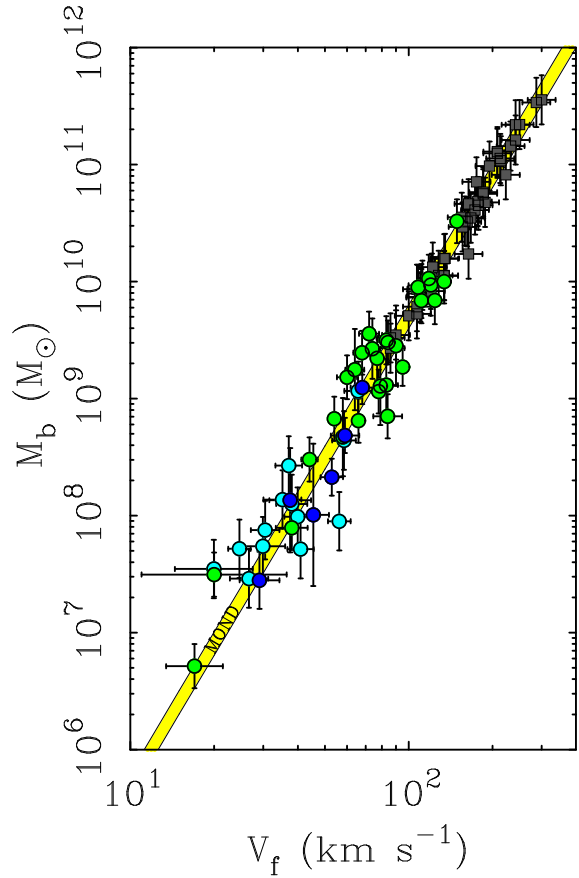}
\caption{The total baryonic mass of disc galaxies plotted against
the plateau (asymptotic) rotational speed. The (yellow) band is the
MOND prediction (whose width represents the uncertainty in $\az$ as
determined by the data). From \cite{mcgaugh10}.}\label{fig1}
\end{center}
\end{figure}
\item
In galaxies where $V^2/R>\az$ in the inner parts, the mass
discrepancy appears always around the radius where  $V^2/R=\az$,
amply vindicated in many analyses.
\item Isothermal spheres (IS)--approximating
ellipticals and other spherical stellar systems--must have mean
surface densities $\bar\Sigma\lesssim \az/G$: Newtonian ISs have,
perforce, infinite mass. In MOND, they have a finite mass, but their
mean acceleration has to be $\av{a}\lesssim\az$ to enjoy the MOND
protection ($\av{a}\sim G\bar\Sigma$). (Non-isothermal spheres, such
as stars, can have arbitrarily high $\bar\Sigma$.) For comparison
with the data see e.g., \cite{scarpa03}
\item
For pressure-supported systems, characterized by a velocity
dispersion $\sigma$, the ratio $\eta=\frac{\sigma^4}{MG\az}$ has to
be of order unity, but depends somewhat on the definition of
$\sigma$, on the degree of isotropy, on the proximity to the MOND
regime, etc.
 This prediction underlays the observed
Faber-Jackson relation between the luminosity and velocity
dispersion of elliptical galaxies, and extends to globular clusters,
galaxy clusters, etc. (e.g., \cite{scarpa03}). Intrinsic scatter, or
systematic deviation, is predicted because of possible scatter, or
systematics, in $\eta$.
\item
The central surface density of ``dark halos'' is $\approx \az/ 2\pi
G$ \cite{milgrom09b,gentile09,donato09}.
\item
Added stability of discs with  $\bar\Sigma\lesssim \az/G$; i.e.,
discs whose bulk is in the MOND regime, where accelerations scale as
$a\propto M\^{1/2}$, instead of the Newtonian $a\propto M$. Thus $\d
a/a=(1/2)\d M/M$, leading to a weaker response to perturbations
\cite{milgrom89a}.
\item  Disc galaxies have both a disc AND a
spherical ``DM'' components \cite{milgrom01}.
\item MOND theories are inherently nonlinear. This
leads to a generic prediction of an ``External-field effect'' (EFE):
The internal dynamics in a subsystem is affected by the acceleration
with which it is falling in a mother system. MOND theories do not
obey the strong equivalence principle, which prevents such effects
in GR.
\end{itemize}

\subsection{Rotation curves of disc galaxies}
At present, full rotation curve (RC) analysis of disc galaxies,
constitutes, arguably, the most comprehensive and robust
manifestation of the mass discrepancy, and offers the best
touchstone for testing alternative dynamics. Given the distribution
of baryons in a galaxy, A MOND MG theory predicts the full
gravitational field of the galaxy. In particular, it predicts the
field in the midplane, which, in turn, determines the rotation
curve. In MI theories, the gravitational field is the standard one,
but particle trajectories are predicted by the specific theory. Even
though we do not yet have a full-fledged MI MOND theory, as already
mentioned, it can be shown that all such theories predict a simple,
universal, algebraic relation between the MOND and Newtonian
accelerations, on circular orbits, in an axisymmetric potential:
relation (\ref{rc}); enough to predict the RC for a given galaxy.
This relation was used in all MOND RC analyses to date. The exact
choice of the universal function $\m(x)$ appearing in eq.(\ref{rc})
turns out to affect only a little the predicted RCs [given that it
satisfies the required asymptotic limits $\m(x\ll 1)\approx x$,
$\m(x\gg 1)\approx 1$]. Small differences between the predictions of
different forms of $\m(x)$ for high-acceleration galaxies, do exist,
of course, and are being explored.\footnote{For the many
low-acceleration galaxies that have been analyzed, where $a\ll\az$
everywhere in the galaxy, only the $x\ll 1$ behavior of $\m(x)$
enters.} Some computations of RC predictions of MG theories have
shown that these do not differ much from those of MI, although
differences do exist, and can, in principle, be used to distinguish
between the two classes of theories.\footnote{some of the
differences may perhaps be compensated by adjusting the choice of
the interpolating functions $\m(x)$, which appears in both classes.
Note that $\m(x)$ appears in different ways in the two theory types.
In existing MG theories $\m$ appears in the gravitational action,
and would enter all phenomena; in MI formulations, $\m$ pertains
only to circular orbits in axisymmetric potentials.} Part of the
reason that the predictions of the two classes of theories are
similar, is that the general form of the RC is predicted by the
basic MOND tenets and so is shared by all MOND
theories.\footnote{The asymptotic velocity is predicted to be
constant and its value determined uniquely from the total mass, and,
the velocity starts from zero at the center. In a high acceleration
galaxy, the predicted velocity takes the Newtonian values roughly up
to the radius, $R$, where $V(R)^2/R=\az$, then the MOND behavior
takes over.} Additionally, for spherical systems, a relation of the
form (\ref{rc}) is also exact for MG formulations.
\par
Contrary to MOND, DM does not predict the RCs of individual galaxies
from the baryon distribution. One can, at best, try to fit a DM
halo, taken from a several-parameter family of DM halos predicted by
this or that concept of DM (for example, the NFW-like halos
predicted by CDM).
\par
\begin{figure}
\begin{center}
\begin{tabular}{lll}
\includegraphics[width=0.36\columnwidth]{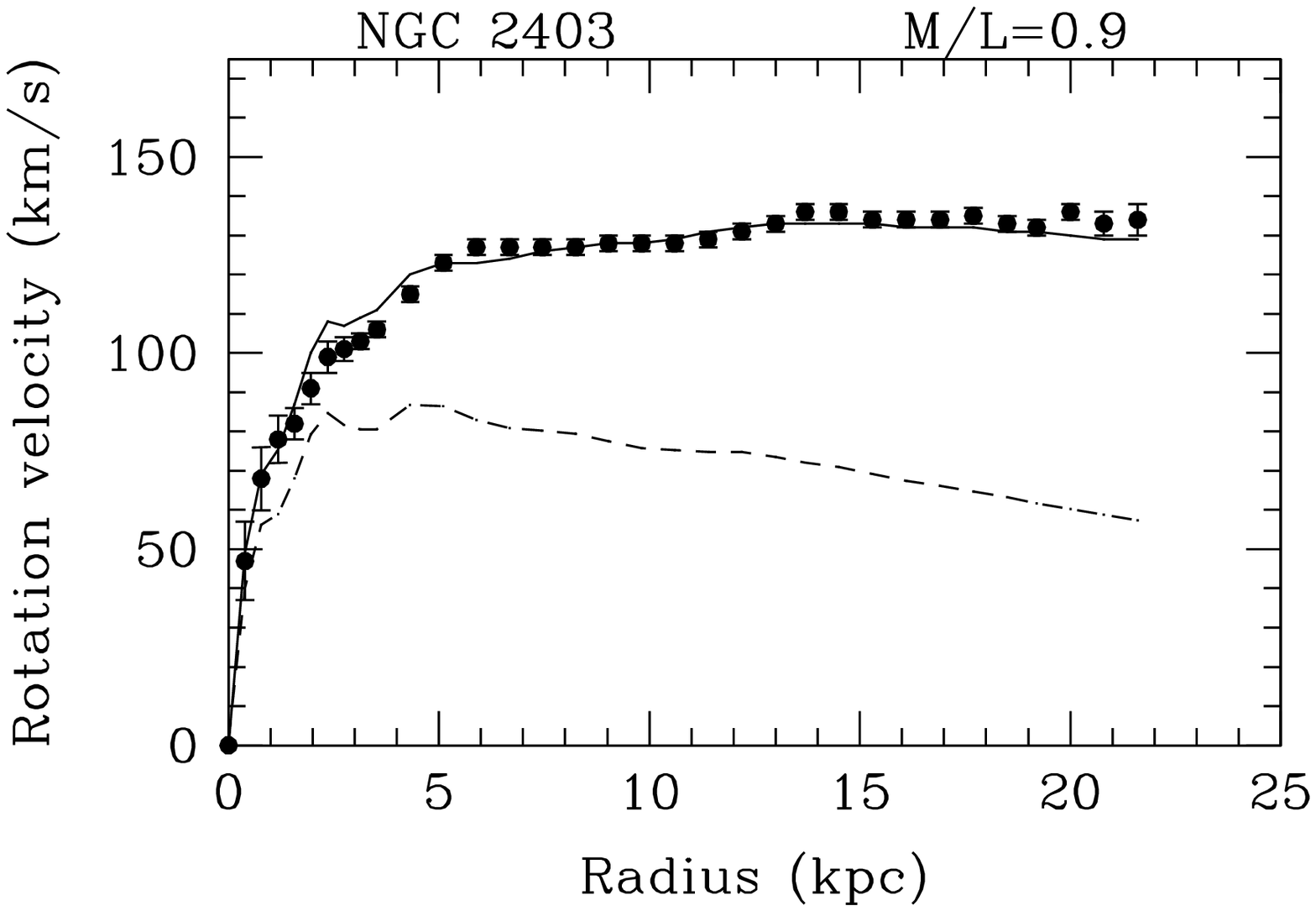}&
\includegraphics[width=0.32\columnwidth]{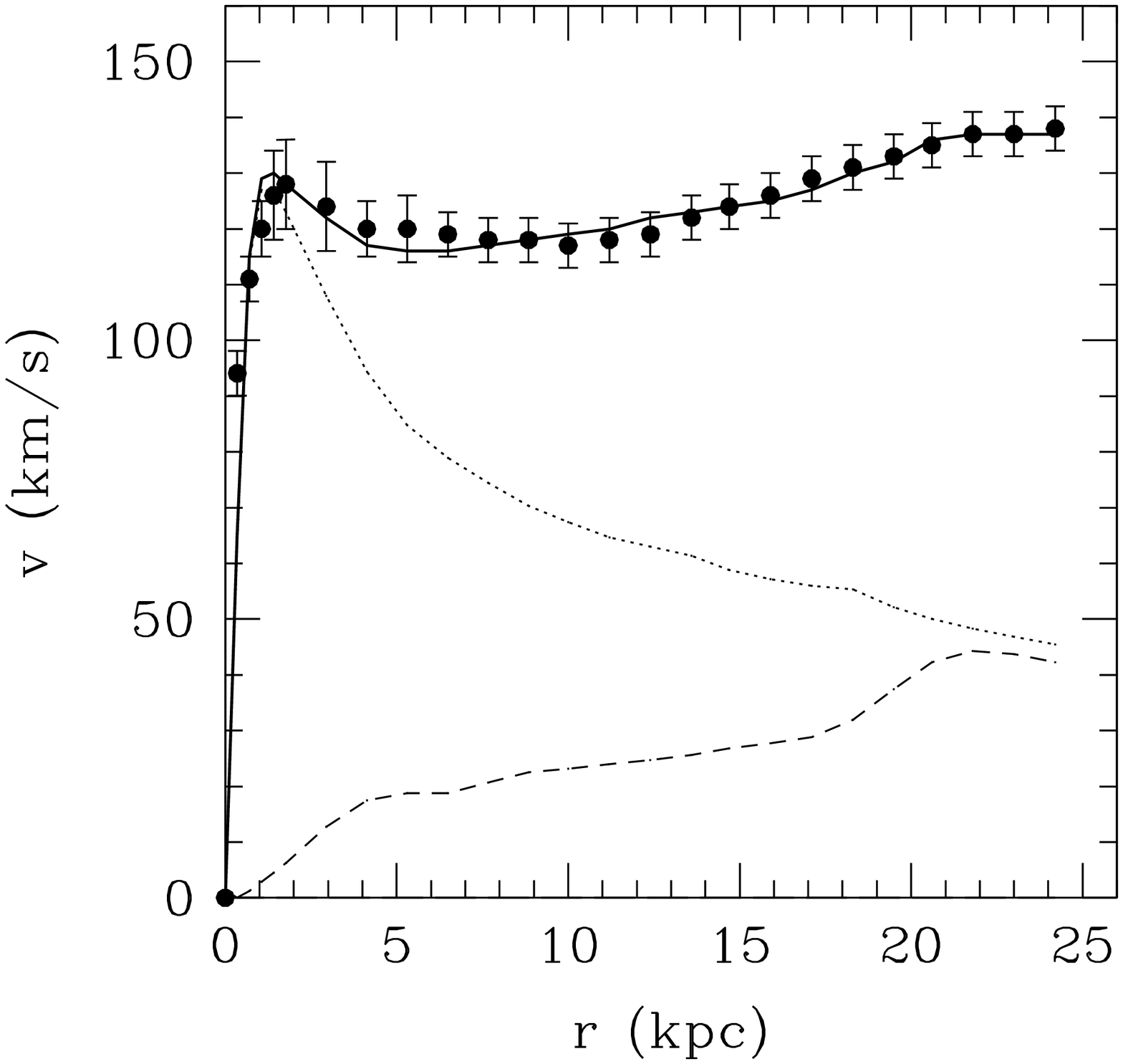}\\
\includegraphics[width=0.65\columnwidth]{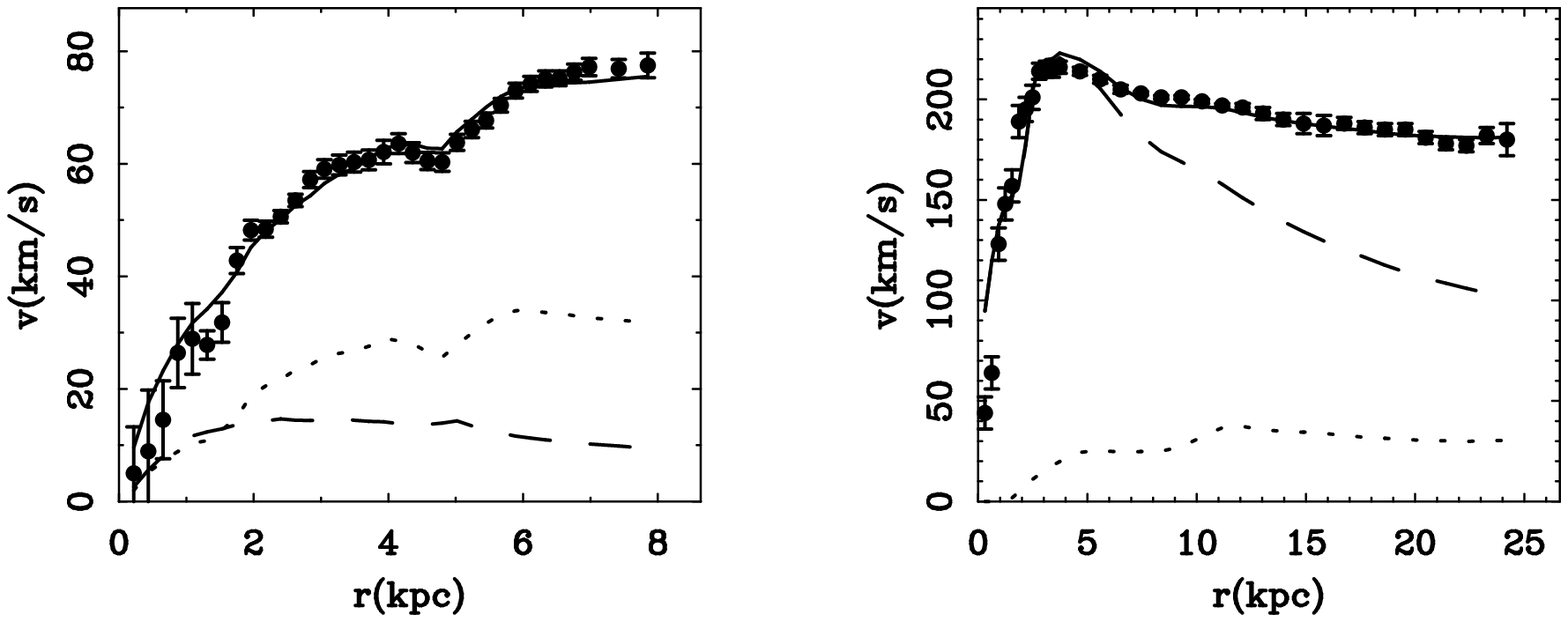}&
\includegraphics[width=0.32\columnwidth]{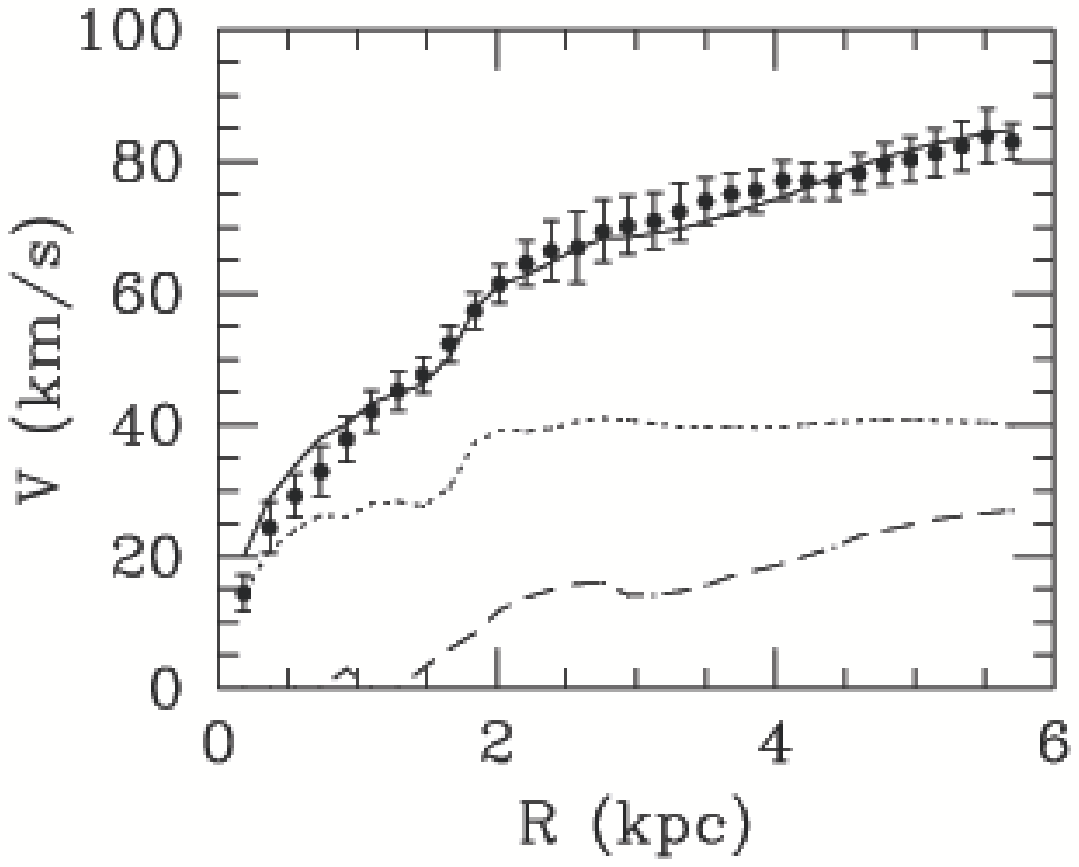}
\end{tabular}
\caption{Rotation curves for galaxies across the gamut of RC types,
from low-mass-low-speed galaxies that are everywhere of
low-acceleration, to high-mass-high-speed galaxies that have high
accelerations in the inner parts. The solid lines are the MOND
curves; other lines are Newtonian contributions from the stars and
gas (from \cite{sm02}, \cite{sanders06}, and curtesy of Bob
Sanders).}\label{fig2}
\end{center}
\end{figure}
\begin{figure}
\begin{center}
\begin{tabular}{ll}
\tabularnewline
\includegraphics[width=0.5\columnwidth]{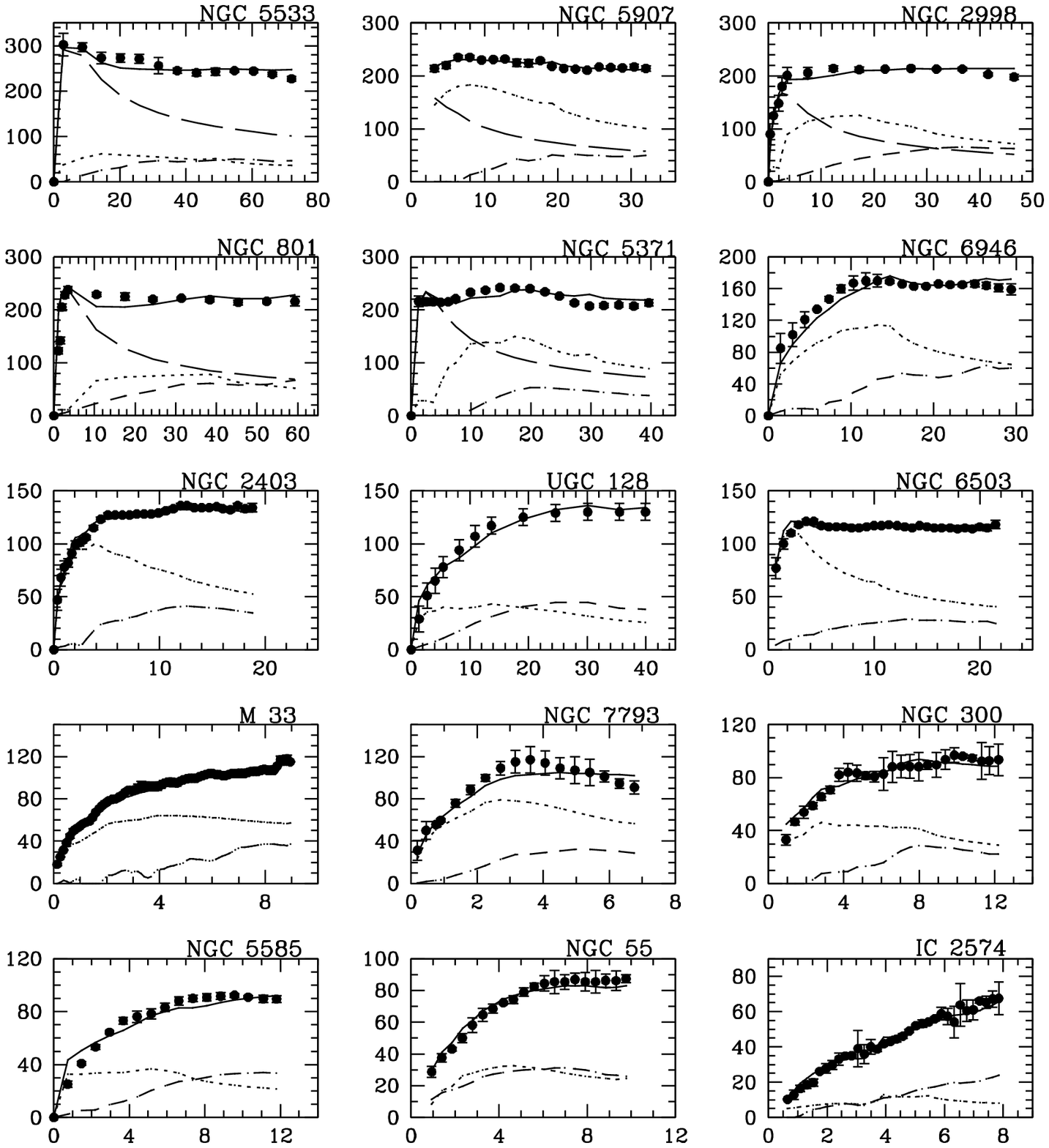} &
\includegraphics[width=0.5\columnwidth]{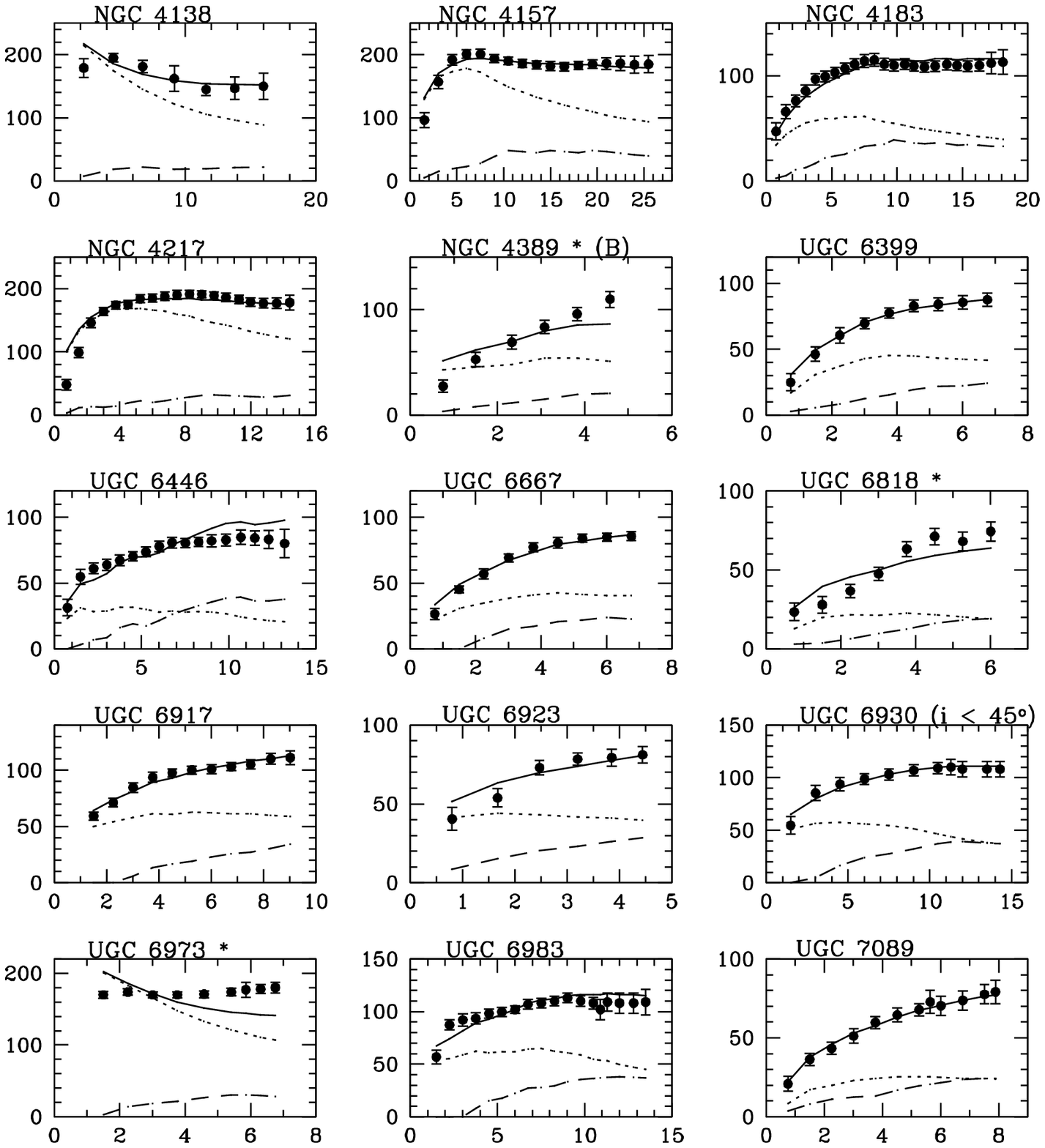}
\end{tabular}
\caption{More MOND results for galaxy RCs (from \cite{sm02}). The
solid lines are the MOND curves; the other curves are based on
Newtonian dynamics: for the stellar disc (dots), gas (dash dots),
and stellar bulge (long dashed) contributions.}\label{fig3}
\end{center}
\end{figure}

Figures. \ref{fig2} and \ref{fig3} show some MOND rotation curves
for disc galaxies. In Fig. \ref{fig2}, I put together galaxies
covering a large part of the range of galaxy types. Some of these
show that the predicted MOND curves reproduce even local features on
the observed RC. These features can be traced back to features in
the baryon mass distribution of the galactic disc. They tend to wash
out even in the best DM fits, since it is not expected that the same
feature that appears in the disc-like baryon distribution, also
appears in the spheroidal DM mass distribution.

\subsection{Elliptical galaxies}
The acceleration field around elliptical galaxies has to be probed
by means other then RCs; these are generically less robust. There
are, e.g., several strong-lensing studies, which probe the very
inner parts; they are inconclusive, as the deduced mass
discrepancies there are small and uncertain. Concentrating on
studies that go to large galactic radii of isolated elliptical
galaxies, I mention here the analyses of \cite{ms03} and
\cite{tiret07}, who find good agreement with MOND prediction, in
analyses based on planetary-nebulae, or companion-galaxy, probes.
Recently, \cite{humphrey10} published a mass analysis of the
isolated elliptical NGC 720, based of x-ray-gas hydrostatics. They
find that the mass discrepancy at a radius of $100 \kpc$ is $\sim
10$. I note here that this is in very good agreement with the
prediction of MOND: The MOND acceleration there is
$g(100\kpc)/\az\equiv x\approx 0.11$, and the predicted discrepancy
is $1/\m(x)$, which is $\approx 1/x$ for small $x$.

\subsection{Systems of galaxies}
MOND has been tested also on systems of galaxies, for example, on
poor galaxy groups, in \cite{milgrom02}, and on super clusters;
e.g., in \cite{milgrom97a}. These analyses showed that the very
large mass discrepancies shown by these systems disappear in MOND,
within the uncertainties.
\par
However, analyses of galaxy clusters, employing x-ray-gas
hydrostatics, and gravitational lensing, shows a persistent,
remaining mass discrepancy, even when MOND is used (see e.g.,
\cite{sanders99,point05}, and references therein, for earlier
findings to this effect, going back to the early 1990s): The typical
mass discrepancy of galaxy clusters within a few megaparces of the
center, which by standard dynamics is of the order of a factor 10,
is reduced in MOND to about a factor of two only. The deduced
density distribution of the MOND ``phantom'' DM is similar to that
of galaxies, and is rather more centrally concentrated than that of
the x-ray gas, which makes the lion share of the baryons. As a
result, the remaining mass discrepancy is rather more pronounced
near the centers of the clusters. A more detailed discussion of this
remaining discrepancy, and of possible explanations of it, are
reviewed in \cite{milgrom08a}, advocating, specifically, that the
discrepancy is due to yet undetected baryons in clusters. Other
suggestions involve neutrinos as the remaining cluster DM
\cite{sanders03,angus09}. The much bruited discrepancy shown by the
colliding, double ``bullet cluster'', is exactly what is expected
from previous analyses of single clusters, and does not further
redound to the cluster conundrum in MOND.
\subsection{Cosmological ``dark matter''}
Accounting fully for all the cosmological effects attributed
conventionally to cosmological DM, is a remaining challenge for
MOND. (The effects of ``dark energy'' can be naturally gotten in
MOND theories, without ``dark energy'', as we saw above, e.g., in
connection with BIMOND.) As has been amply shown, there are MOND
effects that mimic some aspects of cosmological DM (e.g., the
enhanced rate of the growth of perturbations, which occurs in MOND
because gravity is augmented at low accelerations). But a full
accounting may have to await the advent of a fundamental theory of
MOND. I discuss this issue some more in the last section.

\section{$\az$ and its significance}
As shown in sec. \ref{kepler}, $\az$ appears in various laws
predicted by MOND; its value is determined, consistently, by
matching these predictions to the data. Thus, $\az$, and the
associated MOND phenomenology, is here to stay, no matter what the
interpretation of MOND is. Arguably, the best constraints on $\az$
come from the measured proportionality factor of the mass-asymptotic
speed relation (\ref{mvr}), and from more detailed rotation-curve
analysis (for a recent account of such determinations, see
\cite{mcgaugh10}). As shown by eq.(\ref{baza}), the value found
$\az=(1.20\pm 0.25)\times 10^{-8}\cmss$, is near some cosmologically
significant accelerations. Is this a mere coincidence? Assuming that
it is not, and that it does have a fundamental significance, several
questions arise: Is the approximate equality $\baz\approx c H_0$,
where $H_0$ is today's Hubble constant, the more fundamental one,
and the second equality is then just a result of the celebrated and
mysterious ``cosmic coincidence'' $H_0^2\approx \Lambda/3$? In this
case, does this signify that $\az$ varies with cosmic time in
proportion to the expansion rate, so as to maintain the equality at
all times? Then, antropic explanations of the cosmic coincidence
suggest themselves ( \cite{milgrom89,sanders98}). Is the second
equality, $\baz\approx c(\Lambda/3)^{1/2}$, the more fundamental
one?  Is then $\az$, as is purportedly $\Lambda$, a veritable
constant? In either case, the information about the state of the
universe at large, encoded in $H_0$ and/or $\Lambda$, somehow has to
enter local dynamics in small systems, such as galaxies (as
discussed, e.g., in \cite{milgrom99}). Alternatively, we can have a
scheme in which the $\az$ appears in some underlying theory of MOND,
and so it both fixes the CC, and also appears in local dynamics.
This happens, e.g., in the theory discussed in \cite{blt09}, and in
BIMOND \cite{milgrom09}.
\par
The double equality (\ref{baza}) doubles the mystery of the cosmic
coincidence: not only is the expansion rate today comparable with
the (fixed?) value of $(\Lambda/3)^{1/2}$, but both are comparable
with an acceleration constant that appears ubiquitously in the
dynamics of galactic systems.
\par
Another observation that might indicate a connection of the local
physics inherent in MOND with cosmology, is the aforementioned space
conformal invariance of the deep-MOND limit of some NR versions of
MOND. This invariance completes the space symmetry group of the
theory to SO(4,1), which is the geometrical symmetry group of a de
Sitter space-time \cite{milgrom09a}.

\section{MOND vs CDM}
MOND and standard-dynamics-plus-DM are two competing paradigms now
practiced side by side, with the adherents of each complacently
pointing to the successes of their own favorite, and to the
shortcomings of its competitor. The two are, however, not sister
paradigms to be tested on the same touchstones. They account for the
observed mass discrepancies in the universe in completely different
ways, and judging them calls for different criteria. Take, for
instance, the analysis of rotation curves by the two paradigms: It
does not make good sense to judge with the same eye, and on equal
footing, the agreement between the observed and theoretical curves
for the two paradigms, since the MOND curves are inescapable
predictions, while the DM theoretical curves are only fits,
affording great latitude (usually with two parameters).
\par
Consider, specifically, the now prevalent version of DM: the cold
dark matter (CDM) paradigm. According to this picture, the mass
discrepancies observed today in galactic systems are an outcome of
the varied and turbulent histories of individual systems. Such
histories involve the initial collapse of the system, possible
mergers with other systems, violent collisions, gas accretion,
dissipation, baryon heating and cooling, loss of matter from the
system due to supernova explosions, AGN activity, and ram-pressure
stripping, etc., etc. Under these processes, the dissipative and
interactive baryons undergo very different influences from the
presumably weakly interacting CDM. Thus the resulting relations
between the distribution of the CDM and that of the baryons in a
given system--which encapsulates the mass discrepancy--depend very
strongly on the details of the history. These details are not known,
and cannot be known, for a given individual system. CDM can thus
not, generically, predict the mass discrepancy and its distribution,
in an individual object (given the observed baryon distribution).
One interesting and crucial exception to this concerns tidal dwarfs:
the small ``phoenix'' galaxies that are born in gas tails produced
in the aftermath of high-speed galaxy collisions. These dwarfs are
formed in one, relatively clean, process, whereby previous history
is erased. CDM predicts robustly that practically no DM finds its
way into these dwarfs, contrary to what is claimed to be observed
(see, e.g., \cite{bournaud07,metz09}). MOND does predict correctly
the observed mass discrepancies in such systems.
\par
If not predicting the details of the mass discrepancies in
individual systems, can CDM predict, at least, general statistical
correlations between DM and baryons in galactic systems, such as the
different mass-velocity relations? These too would result from the
above mentioned complicated, and little understood, physical
processes, especially those involving the baryons. In CDM, attempts
to predict such general correlations employ so called semi-analytic
models. These purport to account for the complicated physical
processes, putting in by hand all sorts of recipes, and formulas.
But they are all based on guesses that, while educated, are quite
unreliable. In any event, such considerations should necessarily
lead to correlations with very large scattering, as the haphazard
element cannot be avoided--while the observed correlations are known
to be very tight.
\par
What CDM can perhaps do well is predict statistical properties of
pure CDM haloes--inasmuch as they are only little affected by the
small amount of baryons now left in them. For example, the general
statistics of halo population, and properties of the density
distribution within a halo. It is well known, however, that the most
straightforward predictions of this type (regarding, for example,
the number of satellite halos, and the central density cusp
predicted for CDM halos) are in conflict with observations; attempts
to excuse such conflicts require high contortion skills.
\par
In contrast, in MOND, all aspects of the mass discrepancies are
inescapable predictions, independent of the exact system history,
and furthermore they all follow from the distribution of the baryons
alone.
\par
Epistemologically, the situation is similar to that concerning
planetary systems, where we all agree on the underlying dynamics.
Many of the properties of planetary systems at a given time--such as
the number of planets, their masses, compositions, and orbital
radii--result from their complicated, and rather poorly understood,
formation and evolution history. But other properties--such as those
epitomized by Kepler's laws, and the dependence of Kepler's constant
on the mass of the central star--are inescapable consequences of an
undisputed, underlying theory of dynamics. We see in galaxies
phenomenological laws that are as robust as Kepler's laws were, when
they were recognized in the solar system, and which are universal.
Does it make more sense to deduce that such regularities somehow
resulted from a complicated and haphazard formation history of
galaxies, or to accept that they are inevitable consequences of some
appropriate underlying dynamics?
\par
The main challenge that DM advocates now level at the MOND camp--as
also echoed in this conference--is the remaining unproven ability of
a MOND extension to account, as well as CDM does, for some
cosmological observations--such as some aspects of the CMB power
spectrum, and some aspects of large-scale structure. This is true:
what standard dynamics has managed to explain in cosmology by
invoking cosmological DM has not yet been shown to be fully
accounted for by a relativistic MOND version (the effects of ``dark
energy'' can be naturally accounted for in MOND, as I mentioned
above in connection with BIMOND). TeVeS does go some way in this
direction, but not fully, and BIMOND's implications on these points
have not been explored enough, to say nothing of other possible MOND theories.
\par
It has to be remembered, however, that what has already been
accounted for by MOND in galaxies is much more than what still
remains to be achieved in cosmology. In a sense, each galaxy is a
universe of its own, offering as much data to be matched by a
theory, as does cosmology at large. Newtonian dynamics was, after
all, deduced with only one ``mini-universe'' in mind--the solar
system. There is no reason to doubt that some
relativistic version of MOND will be capable of accounting for the
(epistemologically) little that still remains to be explained.


\begin{thebibliography}{}
\bibitem[1]{milgrom83}M. Milgrom, Astrophys. J. 270, 365 (1983)
\bibitem[2]{sm02}R. H. Sanders, and S. S. McGaugh, An. Rev. Astron.
Astrophys. 40, 263 (2002).
\bibitem[3]{bek06}J.D. Bekenstein, Contemp. Phys. 47, 387 (2006)
\bibitem[4]{milgrom08}M. Milgrom, In Proceedings XIX Rencontres de
Blois; arXiv:0801.3133 (2008).
\bibitem[5]{milgrom10}M. Milgrom, In "The Invisible Universe
International Conference", Paris, June 2009 (J.M. Alimi et al.
eds.), arXiv:0912.2678
\bibitem[6]{sanders10}R.H. Sanders, ``The dark matter problem:
a historical perspective'',
Cambridge U. Press (2010)
\bibitem[7]{milgrom89}M. Milgrom, Comm.
Astrophys. J. 13, 215 (1989)
\bibitem[8]{milgrom94}M. Milgrom, Ann. Phys. 229, 384 (1994)
\bibitem[9]{sanders98}R.H. Sanders, MNRAS, 296, 1009 (1998)
\bibitem[10]{milgrom99}M. Milgrom, Phys. Lett. A 253, 273 (1999)
\bibitem[11]{blt09}L. Blanchet, and A.  Le Tiec,
Phys. Rev. D 80, 023524 (2009)
\bibitem[12]{milgrom09a}M. Milgrom, Astrophys. J. 698, 1630
(2009)
\bibitem[13]{milgrom06}M. Milgrom,  in {\it Mass Profiles and Shapes
of Cosmological Structures} G.A. Mamon, F. Combes, C. Deffayet and
B. Fort (eds) EAS Publications Series, 20, 217 (2006)
\bibitem[14]{bm84}J. Bekenstein and
M. Milgrom, Astrophys. J. 286, 7 (1984)
\bibitem[15]{milgrom97}M. Milgrom, Phys. Rev. E 56,
1148 (1997)
\bibitem[16]{milgrom10c}M. Milgrom, Mon. Not. R. Astron. Soc. 403,
 886 (2010)
\bibitem[17]{bek04}J.D. Bekenstein, Phys. Rev.
D 70, 083509 (2004)
\bibitem[18]{sanders97}R.H. Sanders, Astrophys. J. 480, 492 (1997)
\bibitem[19]{skordis09}C. Skordis, Class. Quant. Grav.
26, 143001 (2009)
\bibitem[20]{fs09}P.G. Ferreira, and G.D. Starkman, Science 326, 812
(2009)
\bibitem[21]{sagi9}E. Sagi, Phys. Rev. D 80, 044032 (2009)
\bibitem[22]{dodelson06}S. Dodelson
and M. Liguori, Phys. Rev. Lett. 97, 231301 (2006)
\bibitem[23]{skordisetal06}C. Skordis, D.F. Mota,
P.G. Ferreira, and C. Boehm, Phys. Rev. Lett. 96, 011301 (2006)
\bibitem[24]{skordis06}C. Skordis, Phys. Rev. D 74,
103513 (2006)
\bibitem[25]{skordis08}C. Skordis, Phys. Rev. D 77,
123502 (2008)
\bibitem[26]{zlosnik08}T.G. Zlosnik, P.G. Ferreira, and G.D. Starkman,
Phys. Rev. D 77, 084010, (2008)
\bibitem[27]{sagi10}E. Sagi, Phys. Rev. D 81, 064031 (2010)
\bibitem[28]{zlosnik07}T.G. Zlosnik, P.G. Ferreira, and G.D. Starkman,
Phys. Rev. D 75 044017 (2007)
\bibitem[29]{jm01}T. Jacobson and D. Mattingly, Phys. Rev. D 64,
024028 (2001)
\bibitem[30]{milgrom09}M. Milgrom, Phys. Rev. D 80, 123536  (2009)
\bibitem[31]{milgrom10a}M. Milgrom, Mon. Not. R. Astron. Soc. 405,
1129 (2010)
\bibitem[32]{milgrom10b}M. Milgrom, Phys. Rev. D 82, 043523 (2010)
\bibitem[33]{cz10}T. Clifton and T.G. Zlosnik, Phys. Rev. D 81, 103525
 (2010)
\bibitem[34]{bruneton09}J-P. Bruneton, S.
Liberati, L. Sindoni, and B. Famaey, JCAP, 03, 021 (2009)
\bibitem[35]{zhao08}H.S. Zhao and B. Li, arXiv:0804.1588 (2008)
\bibitem[36]{lc10}X. Li and Z. Chang, arXiv:1005.1169
(2010)
\bibitem[37]{ho10}C.M. Ho, D. Minic, Y.J. Ng, arXiv:1005.3537
(2010)
\bibitem[38]{kt10}V. V. Kiselev and S. A. Timofeev, arXiv:1009.1301
(2010)
\bibitem[39]{rom10}J.M. Romero, R. Bernal-Jaquez, O.
Gonz\'{a}lez-Gaxiola, Mod. Phys. Lett. A, 25 (29), 2501 (2010)
\bibitem[40]{mcgaugh10}S. S. McGaugh, preprint (2010)
\bibitem[41]{scarpa03}R. Scarpa, in
"Structure Evolution and Cosmology", 2002, Santiago, Chile
arXiv:astro-ph/0302445 (2003)
\bibitem[42]{milgrom09b}M. Milgrom, Mon. Not. R. Astron. Soc., 398, 1023
(2009)
\bibitem[43]{gentile09}G. Gentile, B. Famaey, H. S. Zhao, and P. Salucci,
Nature, 461, 627, (2009)
\bibitem[44]{donato09}F. Donato, et al., MNRAS, 397, 1169 (2009)
\bibitem[45]{milgrom89a}M. Milgrom, Astrophys. J., 338, 121 (1989)
\bibitem[46]{milgrom01}M. Milgrom, Mon. Not. R. Astron. Soc., 326, 1261
(2001)
\bibitem[47]{sanders06}R. H. Sanders , Third Aegean Summer School,
``The Invisible Universe: Dark Matter and Dark Energy'',
arXiv:astro-ph/0601431 (2006)
\bibitem[48]{ms03}M. Milgrom, and R.H. Sanders, Astrophys. J.,
599, L25 (2003)
\bibitem[49]{tiret07}O. Tiret,  F. Combes, G.W. Angus, B. Famaey, and H.S.
Zhao, Astron. \& Astrophys., 476 L1 (2007)
\bibitem[50]{humphrey10}P.J. Humphrey, D.A. Buote, C.R. Canizares,
A.C. Fabian, and J.M. Miller, arXiv:1010.6078 (2010)
\bibitem[51]{milgrom02}M. Milgrom, Astrophys. J., 577, L75 (2002)
\bibitem[52]{milgrom97a}M. Milgrom, Astrophys. J., 478, 7 (1997)
\bibitem[53]{sanders99}R.H. Sanders, Astrophys. J. 512 L23 (1999)
\bibitem[54]{point05}E. Pointecouteau and J. Silk,
Mon. Not. R. Astron. Soc., 364, 654 (2005)
\bibitem[55]{milgrom08a}M. Milgrom, New Astron. Rev., 51, 906 (2008)
\bibitem[56]{sanders03}R.H. Sanders, Mon. Not. R. Astron. Soc., 342,
901 (2003)
\bibitem[57]{angus09}G.W. Angus, Mon. Not. R. Astron. Soc., 394, 527 (2009)
\bibitem[58]{bournaud07}F. Bournaud, P.-A. Duc, E.
Brinks, M.  Boquien, M. Amram, U. Lisenfeld, B. S. Koribalski, F.
Walter, and V. Charmandaris, Science 316, 1093 (2007)
\bibitem[59]{metz09}M. Metz, P.  Kroupa, and H. Jerjen,
Mon. Not. R. Astron. Soc. 394, 2223 (2009)

\end{thebibliography}
\end{document}